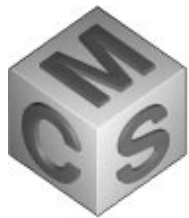

**The Compact Muon Solenoid Experiment**

# CMS Note

Mailing address: CMS CERN, CH-1211 GENEVA 23, Switzerland

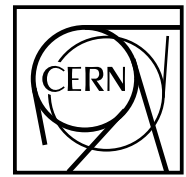

# C.R.I.S.T.A.L.

Concurrent Repository & Information System for Tracking Assembly and production Lifecycles

## A data capture and production management tool for the assembly and construction of the CMS ECAL detector

J-M. Le Goff

*ECP Division, CERN, Geneva, 1211 Switzerland*

J-P. Vialle, A. Bazan, T. Le Flour, S. Lieunard, D. Rousset

*LAPP, IN2P3, Annecy-le-Vieux, France*

R. McClatchey[1], N. Baker[1], H. Heath[2], Z. Kovacs[1]

*1 Dept. Of Computing, Univ West of England, Bristol, UK*

*2 Dept. of Physics, Bristol University, Bristol, UK*

E. Leonardi[3], G. Barone[4], G. Organtini[5]

*3 Univ of Roma I, "La Sapienza" & INFN sec. Rome, Italy*

*4 Univ of Roma III, Rome, Italy*

*5 Univ of Roma III & INFN sec. Rome, Italy*

**Abstract**

The CMS experiment will comprise several very large high resolution detectors for physics. Each detector may be constructed of well over a million parts and will be produced and assembled during the next decade by specialised centres distributed world-wide. Each constituent part of each detector must be accurately measured and tested locally prior to its ultimate assembly and integration in the experimental area at CERN. The CRISTAL project (Concurrent Repository and Information System for Tracking Assembly and production Lifecycles) [1] aims to monitor and control the quality of the production and assembly process to aid in optimising the performance of the physics detectors and to reject unacceptable constituent parts as early as possible in the construction lifecycle. During assembly CRISTAL will capture all the information required for subsequent detector calibration. Distributed instances of Object databases linked via CORBA [2] and with WWW/Java-based query processing are the main technology aspects of CRISTAL. In the prototyping phase, this project will concentrate on the specific needs of the ECAL detector, but intends to address all the CMS detectors with the same complexity.

# 1 Introduction

CMS detectors will be constituted of a very large number of parts. Each element will need to be accurately measured to ensure strict compliance with the nominal quality and resolution of the corresponding detector. Detecting materials, electronics and mechanical structures will be tested and partially assembled in various centres distributed around the world prior to its final assembly at CERN. Much of the information collected during this construction phase will later be needed for the calibration of the detector and to facilitate accurate simulation of its performance. A very reliable software tool is hence required to store the information collected during the detector assembly, to maximise production efficiency and to provide relevant data for calibration and/or simulation.

In the coming years, prior to the final detector construction phase, detector prototyping will be one of the most important activities for CMS collaborators. Performance, resolution and other physical characteristics such as temperature dependence and radiation hardness will be carefully studied for each detector part and the results from test beams will be fed back to the engineering teams responsible for the detector mechanical design and construction. Gradually, the assembly procedures and the measurements required to guarantee the detector quality will be specified. The detector dimensions will be finalised and the construction phase will start. Due to the very large number of specific parts and resources that each detector requires, pre-assembly of sub-parts, tests of the electronics and detector assembly will take place in various specialised sites, called Regional or production centres, distributed world-wide.

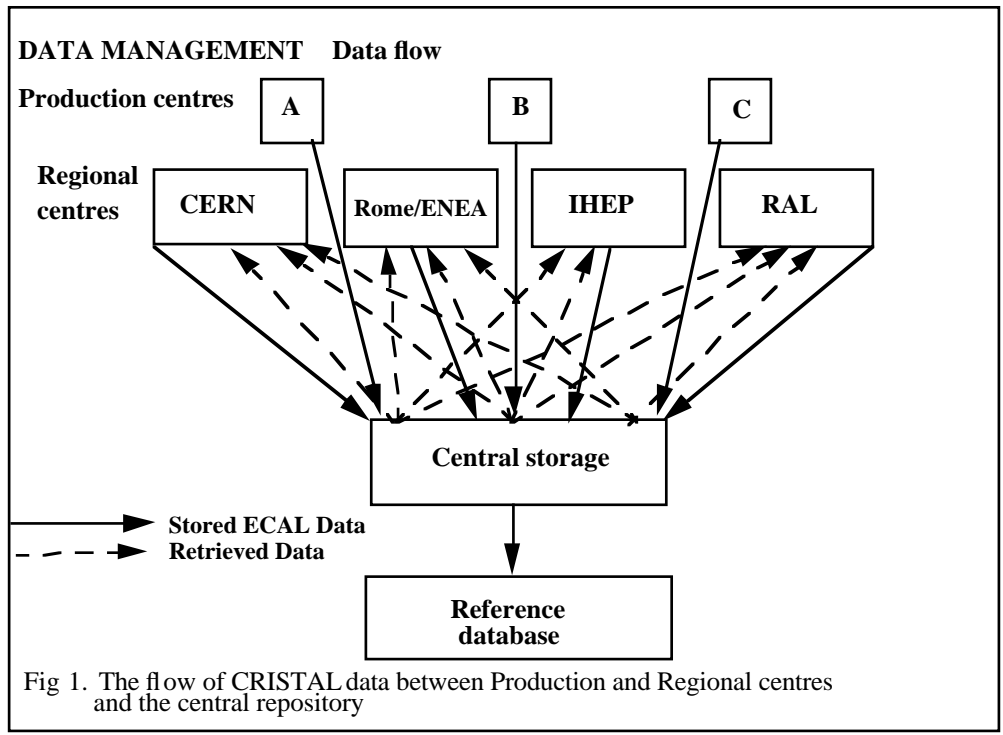

Fig 1. The flow of CRISTAL data between Production and Regional centres and the central repository

To achieve the physics goals, each part produced, must be of the highest quality and carefully checked prior to its final installation in the detector. As an example, consider the CMS ECAL detector. During the construction phase, physical characteristics will be collected and stored by various laboratories for each part of the detector. This information will need to be centralised in a repository and made available for the calibration of the detector (figure 1.)

The production costs must be kept as low as possible. This can only be achieved by a high efficiency in the production. Quality control must therefore be active at each step of the fabrication process. Due to the number of parts, quality control will require computing support. During assembly, matching of detecting elements with electronics may be required to optimise physics performances and minimize read-out channel to channel dispersion. These tasks can be rather complex when production is distributed world-wide and will require computer assistance to the operators of the different centres.

Therefore advanced industrial methods must be used to fulfil the above requirements. This activity includes the quality control of the production, the archiving of the information collected by the instruments and the retrieving of this information for calibration.

After a statement of the CRISTAL project's objectives, this document describes the main features of the software and relates it to other software systems of the experiments such as the Engineering Data Management System [3], the Simulation system, the Calibration and Detector Control Systems. Then this document reviews the main requirements for and concepts behind CRISTAL, and details the technology being considered in the delivery of this software over the next two years. Finally the paper identifies the role of workflow management in CRISTAL and the need for so-called meta-objects in the CRISTAL production management scheme to handle dynamic object schema evolution.

## 2 Project Objective

Overall the CRISTAL development aims to implement a prototype distributed engineering information management and control system that will ensure the production quality in the ECAL assembly programme. This system will be underpinned through the use of a Object Oriented Database Management System (OODBMS) and a set of interfaces accessed via the CORBA [2] standard (see section 6). The system shall provide secure access to the production data and enable the migration of these data between geographically separated production and assembly centres. Its specific objectives are:

• to investigate aspects of information and workflow management for a distributed production line

• to integrate distributed data management based on dynamic WWW applications

• to investigate multiuser and multirole customised access to this engineering information system (EIS)

• to produce a CORBAcompliant prototype of a distributed workflow management system

• to provide rescheduling and redirection facilities of the data flow in the staged production line process from a remote management centre (on request)

• to support the migration of data between stages of the production line

• to provide controlled and secure access (for information retrieval and information presentation) to the distributed engineering data

The current project, entitled CRISTAL (Concurrent Repository and Information System for Tracking Assembly Lifecycles) will facilitate these project objectives. In the first instance CRISTAL will manage the production process of the 110,000 lead tungstate (PbWO4) mono-crystals, and their associated fast electronics, to be installed in the CMS Electromagnetic Calorimeter (ECAL). The software produced will be generic in design and will therefore be reusable for other CMS detector groups.

## 3 CRISTAL users

Seven different types of users of CRISTAL Prototype users have been identified: the producer, the centre operator, the centre supervisor, the physicist writing analysis and reconstruction programs, the technical coordinator, the engineering information system and data storage manager and the instrument data provider

The producer works in a production centre and registers new parts (for example, newly grown crystals) in the CRISTAL Prototype 2 system. S/he (hereafter he) interacts with the system to start/stop a session, to browse the characteristics of the parts created in his centre and to execute registration tasks. Producers are normally commercial oufits which have been contracted to produce the physical detector parts.

Centre operators are located in the various local centres (production and assembly)., will perform all the operations (wrapping, gluing) on the crystals or on any detector parts and will operate all the instruments which provide the physical data. In addition, they will capture manually all the parts information which will not be automatically collected by the instruments. Operators are technicians with no specific knowledge in computer technology. They are expected to use the CRISTAL Prototype 2 system to capture the information relevant to the production process and to operate computer controlled instruments which will populate the data storage automatically. About 10 to 12 operators per centre are expected to use CRISTAL Prototype 2 interactively. An operator will interact with the system to start/stop a session, to move sessions around the computers used to interact with the system, to browse part characteristics and to execute tasks on parts.

There will be one centre supervisor per local centre (production and assembly). The centre supervisor is responsible for the quality control of the crystals or detector parts locally produced or assembled. His main task is to accept or reject the various parts and to monitor and optimise the detector assembly. He is an engineer or a physicist with a background in High Energy Physics and/or in scintillating material sciences. He is expected to browse the engineering information system to supervise the process and to interact with CRISTAL Prototype 2 on a daily basis. Being the supervisor of a local centre, he will be able to execute any of the tasks that an operator can execute.

Physicists are users interested in obtaining information about a detector of the CMS experiment. They will perform two distinct activities. On the one hand, they will consult the detector reference database using standard browsing facilities. On the other hand, they will write software programs to perform analysis tasks using the data contained in the detector reference database.

The Technical Coordinator is responsible for the full ECAL construction lifecycle. He is physicist with a background in High Energy Physics and in scintillating material sciences. He is expected to provide the ECAL quality control technical specifications and to browse the engineering information system to supervise the process. In particular, he is responsible for the definition of the production procedures, the workpackage steps and for the specifications of data capture operations related to the various measurements. The coordinator will interact with the system to define a local centre, to create production scheme elements (parts, characteristics, fields (data types), tasks, batches) to handle versions of the production scheme (a complete set of chained tasks which will control the production process), to handle requests of acceptance/rejection on parts, to handle ordering and shipment of batches of parts and to browse the central storage to monitor production

The engineering information system and data storage manager is the user responsible for the maintenance of the CRISTAL Prototype 2 system in a centre (local or central). He is a physicist or an engineer with a background in computing. His main responsibility is to make sure that the system is fully operational and available in the various centres. He also verifies that no information is lost when being migrated from one centre to another. Each centre (production and regional) will have such a user. In addition, in the central system, this user will be in charge of the maintenance of the detector reference database. The engineering information system and data storage manager will interact with the system to setup a CRISTAL Prototype 2 system, to setup a local centre to manage local storage and to handle system upgrades.

The instrument data provider is responsible for the integration of a physical device or a piece of software into the CRISTAL Prototype 2 system. He is an engineer or a physicist with a background in computing. He is expected to provide the software which will be automatically activated upon task execution and which will provide CRISTAL Prototype 2 with the characteristics (physical data) that will need to be stored. The instrument data provider will interact with the system to define an instrument and to declare a piece of software.

# 4 Relation with other software activities

CRISTAL will need to be connected to other software systems employed in either the construction of the experiment itself or in its operation. The various links are illustrated in figure.2 and are described in the following sections

## 4.1 Engineering Data Management System

The EDMS (Engineering Data/Document Management System) [3] software is defined in the technical specification annex 1: IT-2374/PPE issue 1, revision 0 "The EDMS will be used to manage, store and control all the information relevant for the conception, construction, and exploitation of LHC accelerator and experiments during their whole life-cycle, more than 20 years". In particular all the engineering drawings describing the various detector components and the description of all the engineering assembly procedures will be stored in EDMS. Clearly CRISTAL will need a subset of this information (such as part dimensions and tolerances) to facilitate control of the measurement tasks in production and assembly of the CMS detectors and to match the desired mechanical parameters with the actual measurables.

The main difference between EDMS and CRISTAL is that an EDMS (commercially referred to as a Product Data Management tool) addresses the engineering data aspects of the detector production (e.g. CAD/CAM drawings and product descriptions), whereas CRISTAL addresses the capture and management of (versions of) actual sets of measured characteristics (e.g. transmission spectra, light yields, attenuation lengths) and the allocation of the measuring tasks (and parts) to operators and instruments in (measuring) centres. An EDMS is therefore document management-oriented and handles the mechanical specifications of the detector or experiment, whereas CRISTAL covers aspects related to the control of and automatic data collection from physical instruments, and the corresponding quality control required during detector production and is therefore workflow management-oriented.

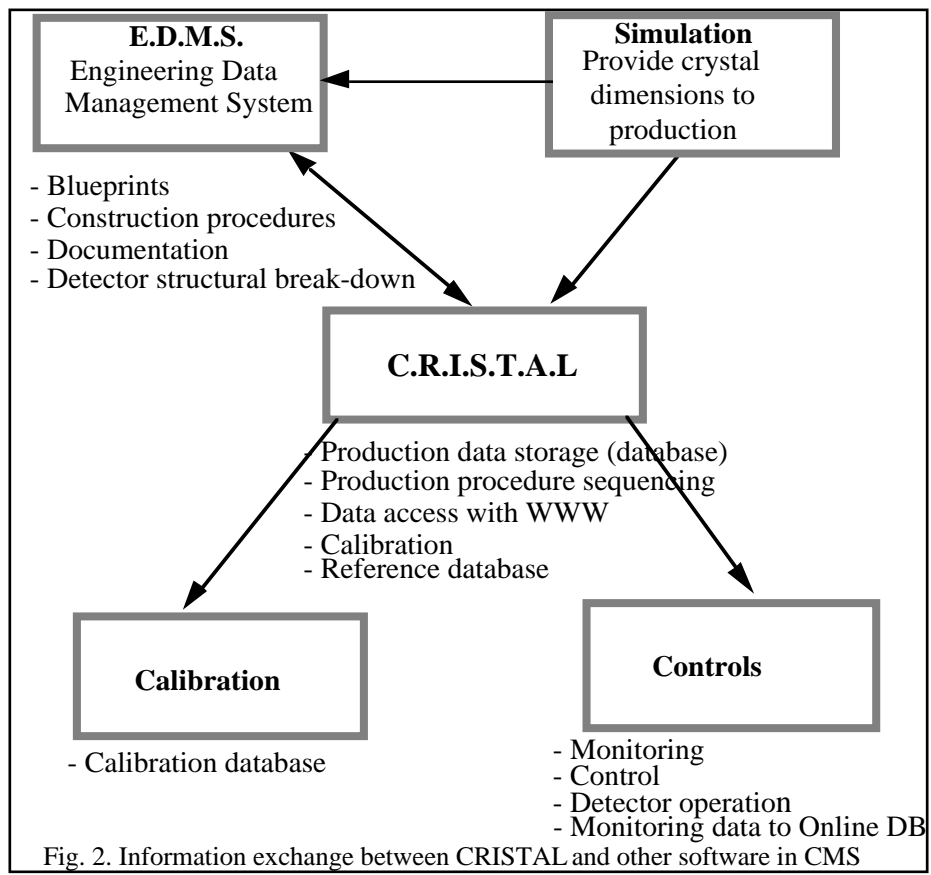

Fig. 2. Information exchange between CRISTAL and other software in CMS

Ultimately, CRISTAL may be required to provide EDMS with some of these physics parameters for future upgrades of the experiment. Hence, CRISTAL is both a client of and a server for data from EDMS and an interface will need to be specified for this sharing of data.

## 4.2 Simulation

Detector simulation software will provide CRISTAL or EDMS with the ideal parameters (such as physical dimensions of detector elements) for all the detecting devices according to an optimised geometry for the experiment. The mechanical operations applied to the raw material like machining, grinding or polishing will be adjusted to provide detecting devices with dimensions and performances as close as possible to the ideal values supplied from the simulation software. During the production and assembly process, CRISTAL will therefore be required to store all the deviations from the ideal specifications in its data storage. These deviations will be provided for the calibration storage to be used for event reconstruction and detector control (see next section).

## 4.3 Calibration

The storage system for calibration software will use a subset of the CRISTAL information for event reconstruction. The data representation in the repository will be very different since the calibration software is mostly servicing software programs and should reflect the structure of the detector from the standpoint of the data acquisition system whereas CRISTAL is interfaced to WWW users and should reflect the structure of the production scheme. For calibration the storage required is mostly static offering fast access to its content. CRISTAL is a production and assembly sequencer which, among other activities, will populate a static storage. As illustrated in the above figure, the link between CRISTAL and calibration is one way.

### 4.4 Controls

During the assembly phase, various probes such as temperature sensors will be calibrated and installed. CRISTAL will collect the probe characteristics and locations. Once installed, most of these probes will no longer be physically accessible. If required, replacement will only be possible during long shut-down when the detector will be removed from its operating position for repair. Therefore any information relevant to the control system must be collected during the assembly phase. These information will be provided by CRISTAL to the Detector Control Systems (DCS) as static data for the read-out of the monitoring systems.

The DCS may diagnose some control probes as having non-optimal behaviour and shall inform CRISTAL to mark them as such in its storage. As a consequence, the DCS will be one of the clients of CRISTAL.

## 5 Requirements in CRISTAL

### 5.1 OODB requirements

As a result of completing the first phase of CRISTAL, the user requirements of the project have been defined. These requirements have been studied to ensure that they can be satisfied using OODB technology. The essential object oriented features that any database used in CRISTAL must provide include the standard OO features of Object Identity, inheritance, complex object handling and secure message passing. In addition the minimum DBMS features that CRISTAL requires of an OODB include persistence of objects (by attachment (deep persistence) or by direct persistence), controlled access to objects, effective querying and browsing, a multi-user capability and full recovery facilities.

CRISTAL requires the storage of multi-media data, so that the object database must be able to handle large documents, pictures, drawings, source code etc. It should provide mechanisms to handle data clustering over multiple distributed sites and manage TeraByte-sized bases. Given the evolving nature of CRISTAL, the database should handle object versioning. Although raw performance of the database is not a limiting issue in CRISTAL, there should be minimal performance losses seen when the number of users increases or when the database size grows. The OODBMS needs to cope with ACID transactions with implicit locks as well as long term transactions (i.e. transactions with explicit locks).

ODMG-93 [4] compliance is a necessity in CRISTAL since choice of technology needs to be deferred until later in the project. This will facilitate change should that be required over the considerable timescales of the CRISTAL project. ODMG compliance requires the provision of language bindings (C++, Java, etc.) a standard Object Query Language (OQL), direct persistence, locking and the ability to move applications from one database to another.

The ODBMS must provide a set of tools (preferably graphic in nature) to manage the database, to support distributed user developments and to facilitate rapid application development. The ODBMS must allow both programmes and object methods to be stored in the database and should provide schema evolution facilities. It should be easy to build the data model from a Unified Method schema definition. The database management system must also provide administration facilities to grant access to users, to handle backups and all the activities related to the management of a multi-user database. The database is required to cope with recovery after a crash and should (ideally) provide roll-forward from a backup to a point in time.

The OODBMS must be accessible from geographically remote (WAN) sites connected by different distributed systems software. These should include CORBA, IDL / ODL, OLE / COM and WWW access. It should also provide server hardware heterogeneity. Heterogeneous distribution is crucial for CRISTAL, but can only be provided by an OODB if the underlying language is interpreted and hence able to be executed by any machine. In the first phase of prototyping CRISTAL has used the O2 [5] product. Presently O2 is based on compiled C++ or C and therefore the code is not directly portable from one machine to another. This problem will be overcome in later prototypes with the use of Java code ultimately being stored as O2 objects.

All the OODB vendors are presently developing query languages, usually based on SQL. In CRISTAL, objects use navigation to search for related objects, either by referencing sub-objects or by holding specific relationships. The O2 query language has been identified as providing the functionality required in CRISTAL.

### 5.2 Production Management

Facilities are required in CRISTAL to monitor the ongoing production processes and thereby to establish the state of the overall detector production process given the activity in the distributed centres. Furthermore it will be necessary to identify potential problem areas in the overall production process (such as bottlenecks), to perform

'what-if' scenarios given the existing production schemes and to recommend to the Coordinator actions to correct deficiencies in the production/testing/assembly process. Such actions may require redirection of the allocation (and/or sequence) of tasks between the centres and may result in new versions of the production scheme. CRISTAL must therefore be able to cope with multiple versions of part definitions, task assignments and production schemes and allow seamless navigation between these parts and schemes. In addition, the Coordinator will be provided with facilities to place orders for parts to centres and to reschedule tasks between centres when it is deemed appropriate. Petri-Net [6] notation is being investigated in CRISTAL to model the overall production schemes (see later section on 'Workflow Management and Meta-Objects in CRISTAL').

### 5.3 CRISTAL Interface with Instruments

The CRISTAL project also aims to provide facilities to assist operators in executing tasks on parts. The knowledge of the state of the part will be held in CRISTAL as will the definition of the tasks required to be carried out by an instrument. Then operators will be assisted in the completion of each production scheme by the part inquiring of the CRISTAL repository the next task to be carried out by a particular instrument. CRISTAL itself will instruct the instrument on which characteristics are to be measured. In addition the project will provide the facility to interface any measuring instrument to CRISTAL via a simple ASCII-based TCP/IP protocol. Users will be able to supplement the CRISTAL system by supplying equipment-specific software to tasks (for example adding inventory checks prior to shipping tasks being executed). This software can then be triggered by tasks in CRISTAL and can take measurements for storage in the database.

## 6 Technology Proposed for Use in CRISTAL

### 6.1 Object-Oriented Development Methods and Tools

Since the timescales for the use of CRISTAL are considerable, care must be exercised in the choice of technology in designing the management system. Object-oriented techniques are being used to develop CRISTAL, to provide reusability of software and to maximise reliability. In order to understand and refine the user requirements for CRISTAL a series of prototypes will be developed. The development of CRISTAL will therefore follow a Spiral Model [7] or evolutionary approach which allows for the planned delivery of multiple releases of the software.

As the development of CRISTAL involves a geographically distributed software engineering effort and since much of the implementation will be carried out by developers on short-term contracts at CERN, there is a real need for clearly defined deliverables and interfaces between (sub-) systems. The use of the European Space Agency Software Engineering Standards (PSS-05) [8] s thus an essential feature in CRISTAL. PSS-05 is being used in harness with an OMT/Booch (Unified Method) [9] approach to the software specification in CRISTAL, supported by the Software Through Pictures (STP) CASE tool.

### 6.2 The CORBA Standard

The Object Management Group (OMG) is an industry consortium dedicated to creating object management standards necessary to achieve the goal of interoperability between heterogeneous, distributed object based systems. Object Management Group's Common Object Request Broker Architecture (CORBA) standard provides the so-called object request broker (ORB) for heterogeneous distributed systems such as will be needed in CRISTAL. The ORB provides the mechanisms by which software and data modules, or objects, communicate transparently across distributed systems. The ORB provides interoperability between applications on different machines in distributed environments and seamlessly interconnects multiple object systems. The CRISTAL project will implement a prototype engineering information system and a set of CORBAcompliant interfaces for the data management, analysis and control programs required for CMS construction (see figure 3). The CRISTAL prototype will be developed using the IONA Technologies implementation of CORBA, ORBIX.

### 6.3 Object Oriented Databases

Since an object-oriented approach is being adopted in the development of CRISTAL and since conformance to emerging software standards is essential to ensure the viability of software developed to be used over a long period of time, object-oriented database (OODBMS) technology has been proposed as the repository technology in CRISTAL. The Object Database Management Group (ODMG) has put forward a set of standards allowing an OODBMS user to write portable applications. This proposed standard should ensure interoperability between different OODBMS products, thereby facilitating the development of a (potentially distributed) heterogeneous database system (communicating via the OMG Object Request Broker) for the capture of production, testing and assembly data in the construction of CMS detectors. For the purposes of prototype development and technology evaluation, the CRISTAL development will be carried out with an ODMG-compliant [4] product as its underlying OODBMS.

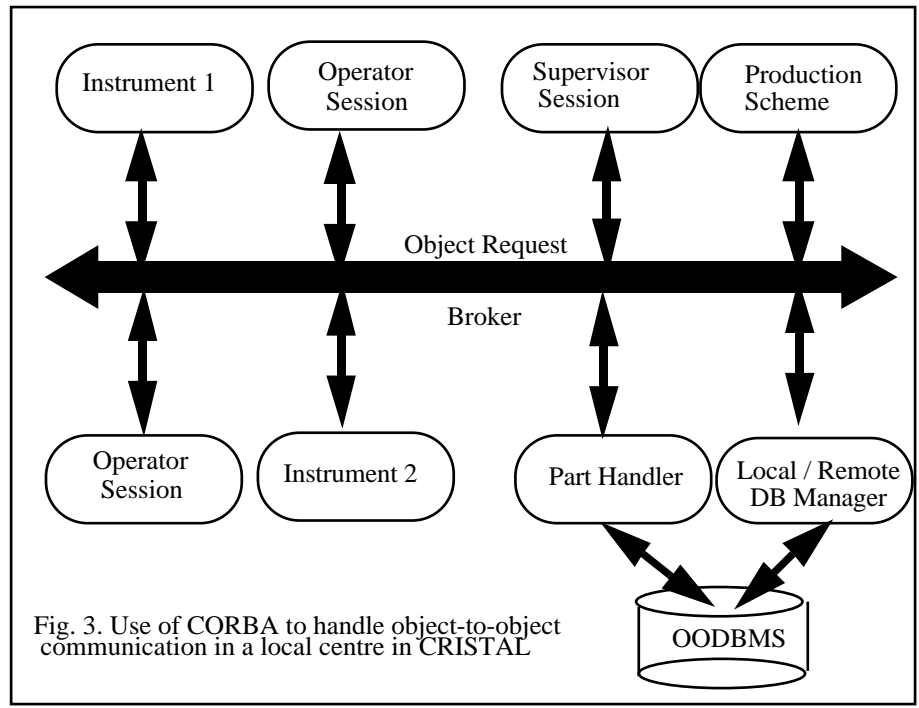

Fig. 3. Use of CORBA to handle object-to-object communication in a local centre in CRISTAL

### 6.4 User Interfaces to the CRISTAL System

The CRISTAL system must provide access to OODBMS-resident data for operators of equipment and for the instruments used to perform the extensive range of tests that are required to measure the physical characteristics of the detector materials. Additional access is required both at the centre level and over the set of centres to ensure quality control in the production and assembly of the detectors. Database managers will also require specialised access to the CRISTAL repository.

The accumulated CRISTAL data will be made available for physicists and engineers for the assembly and calibration of the CMS detectors and ultimately for use in the simulation programmes. All access will be provided via the World Wide Web through (Java-based) programmes for each category of CRISTAL users. The information which is accessed will be filtered according to the role of the users. Java offers a powerful basis for developing such platform-independent applications.

The integration of the Java interfaces with the CORBA environment will be facilitated via IONA's OrbixWeb product. This product enables the Java interface clients to interoperate seamlessly with the back-end CORBA services over a heterogeneous system of machines.

## 7 Research in CRISTAL

### 7.1 Concept Definitions

The philosophy adopted in CRISTAL is to provide the production control and assembly management facilities required by the CMS detector groups in as flexible, open and user-friendly environment as possible. The CRISTAL software will be able to support the concepts of parts (with part definitions) for production and assembly and tasks (plus definitions) to operate on those parts. Parts can be defined to be any atomic pieces of equipment or may be defined to be collections of parts which have relevance for testing or assembly (so-called superparts). Parts are allocated identifiers and can be used for navigation through the construction hierarchy of the detector. Tasks are the elementary stages of the production/testing/assembly lifecycle which have been clearly defined by the Coordinator, which are triggered by human intervention and which behave according to a specification provided by the mechanical engineers. Information (characteristics, measurements) will be captured in CRISTAL as a consequence of the execution of a task on a part. One clear example of this is the measurement of some physical characteristic (such as Light Yield) which is recorded in the CRISTAL database for later use in calibration software. The process Coordinator will define a production scheme (or a sequence of correlated or uncorrelated tasks) which determines the order of tasks which must be applied to the parts. Figure 4 shows a subset of a production scheme for one part definition in centre CTR_05. Five tasks are shown for the part, some of which are strictly sequential in nature (T0, T5), some simple alternatives (T1, T2, T3), where order of task execution by operators is unimportant, and some correlated which must be completed when another particular task has been executed (e.g. T4 correlated to T3). Any modifications to part or task definitions (or task locations) must be reflected as soon as possible in a change to the production scheme to ensure consistency in the production control and assembly process. All such modifications must be recorded in the database, thereby creating traceability of parts and a full historical record of part definition changes. The part itself is responsible for determining from the production scheme which task must be executed next on itself.

Over time the part and task definitions will evolve as a result of knowledge that emerges during detector testing and construction. Therefore, one very important aspect of CRISTAL is the ability to track such changes in the database, to handle versions of these definitions and for (versions of) production schemes to handle the transitions from one version to the next. The Coordinator of the detector construction will specify and modify the part/task definitions in CRISTAL and CRISTAL itself will handle the coexistence of multiple versions and the correlations between these versions.

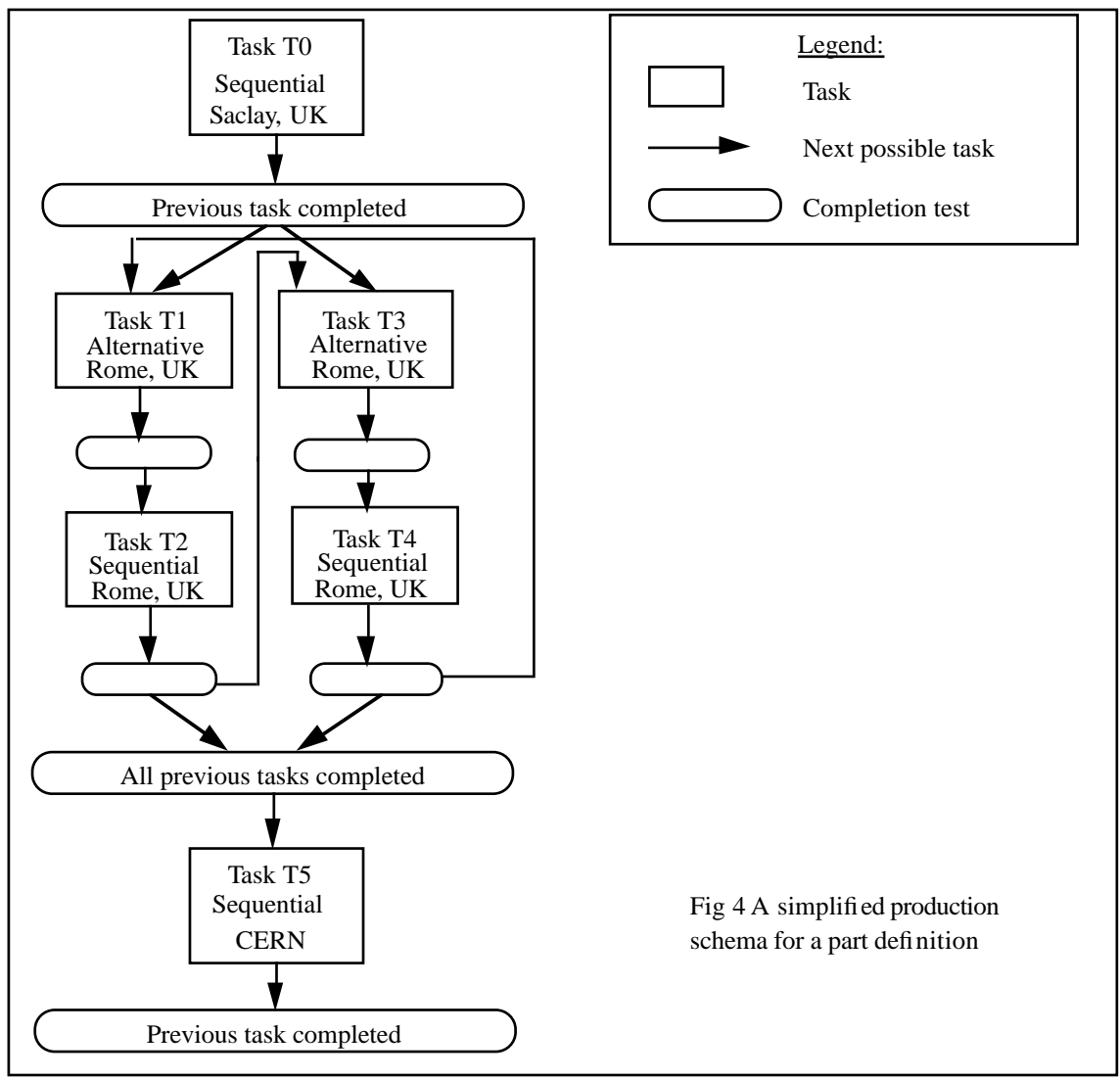

Fig 4 A simplified production schema for a part definition

## 7.2 Object Storage in CRISTAL

The data handling and storage aspects of CRISTAL must be transparent to any users of CRISTAL (operators, centre supervisors, physicists, Coordinators). Part, task and production scheme data will be defined by the coordinators and captured in the database. The coordinators will determine when those definitions become active in the centres and the CRISTAL software will distribute these definitions to centres and thereafter populate the new structures in the centres. Since parts (potentially of common part definitions) may be produced in several centres, measured and tested in other centres and assembled in further centres, it will be necessary for data to be accumulated for a part across distributed centres. Objects in the database must therefore be moved from one database to another in a secure fashion and queries may often need to be executed over several database sites. Consequently, CRISTAL must handle the definition of shipping and reception tasks when parts are moved between centres. In addition, the CRISTAL system must handle all aspects of object migration between database instances and will facilitate the direction of queries to appropriate data stores.

Data will be collected at each centre during the production/assembly lifecycle and will reside at that centre for as long as the part resides at that centre. Production data will be collected at Production Centres while data originating from physics tests and measurements or as a result of assembly will be collected at Regional Centres. Data is collected locally since each centre must be able to operate autonomously thereby avoiding network dependencies. Data collected locally will be duplicated in the central system for security, for access from other centres and to act as input to the final experimental Reference Database. Therefore an object duplication strategy is required. Since the projected final amount of data in CRISTAL is of the order of 1 Tera Byte, a federated database approach with 10 centres (and therefore 10 Tera Bytes of data) is not feasible. The object duplication strategy being formulated in CRISTAL is designed to only keep data in local databases for parts held locally.

It is envisaged that (particularly in the early stage of construction) there will be many changes to the data structures in CRISTAL. If the definitions of the parts and tasks are kept statically in the database then problems of database schema evolution may result when definitions are moved from one database instance to another. For example, a change in the detector setup (such as a change in (super)part definition) may require changes to the object database schema when the part is moved from one centre to another. To circumvent this problem, CRISTAL uses the concept of meta-objects or descriptions of objects which are managed by the database. Meta-objects are customisable without affecting the underlying database schema. An example of a meta-object is a characteristic (such as transmission) whose definition may change from one part to another. The meta-objects as defined in CRISTAL perform two main functions: firstly they handle any amendments to the structure of objects (parts, tasks etc.) without changes to the underlying data schema and secondly, as CORBA objects retrieved from the object database, they provide the dynamic aspects of interaction between the users and the database. For example, the meta-object for parts are accessed via the Part Handler (see fig. 3) which communicates with user sessions via the Object Request Broker.

## 7.4 Workflow Management and Meta-Objects in CRISTAL

From the standpoint of the overall Coordinator, the CRISTAL system can be considered as a workflow management system [10] in that it keeps track of the activity of the production system in a database and stores a history of (the status of) all tasks executed on all parts. In addition to this, CRISTAL exhibits further workflow management characteristics in that it must handle the frequent rescheduling and/or redirection in the flow of work between the Production and Regional Centres in as flexible a fashion as possible. Bonner et al. [11] showed that in such production workflow systems dynamic schema evolution should be supported, this being particularly true when the system supports a high-throughput and/or when the system must provide flexibility in a rapidly evolving workflow environment. They advocate the use of a Laboratory Information Management System [12], called LabBase, based on an OODB (ObjectStore) in their genome laboratory work. Bonner et al. note that LabBase must support both the concepts of event histories (audit trail of workflow activity) and dynamic schema evolution. The CRISTAL system will also need to capture event histories to allow the Coordinator to report on the production activities and to investigate the cause of production bottlenecks amongst other daily production management activities.

CRISTAL therefore shares some similarities with commercial workflow management systems. However, as a result of the severe constraints on the CRISTAL research and development program (time, people, costs), the workflow management problems that arise show differences from those in commercial examples of production workflow systems [13]. Firstly, the nature of the construction of CMS means that the production line will result in exactly one complete product (the detector) at a fixed point in time, and this 'once-off' production must be complete and correct at that point. This implies quite different constraints than those normally experienced in workflow management environments such as car part production, insurance claim handling, telecoms service order provisioning etc. CRISTAL will also be distributed widely over continents (China, UK, Russia, Italy,

CERN, etc.). This is in contrast to production workflow environments which are normally distributed (over short distances) only to facilitate decentralisation of the production. The CRISTAL Coordinator can also redirect the flow of parts between the Production and Regional Centres when it is deemed necessary and can redefine new production schemas with resequenced tasks or can reassign tasks between the Centres. CRISTAL is also required to catalogue versions of the production scheme elements such as task definitions, parts definitions and production scheme definitions. Commercial workflow packages such as ICL's TeamWARE, IBM's Flow Mark or XSoft from Xerox normally provide only limited capability for redefining workflows and operations on those flows. In addition, CRISTAL requires that the workflow management software be interfaced with an OODBMS - another aspect uncommon in commercial applications.

Other areas which make CRISTAL somewhat unique are the fact that the scale of the data storage is vast (TeraBytes) whereas the number of users is actually small (< 100), the fact that CRISTAL requires storage and comparisons of actual physical measurements rather than simple production statistics, the fact that CRISTAL requires real database administration facilities such as backup and recovery, and the fact that the production line will run and evolve over such an extended period of time (at least 6 years). In addition to these, task management in CRISTAL is complex since tasks can be repeated or undone at any time by the operators: these activities cannot be handled by current commercial offerings. As a result current commercial workflow management products are inappropriate for the CRISTAL development (although research work has been initiated elsewhere into expanding commercial databases to provide workflow management capabilities [14]) and we must 'tailor-make' the production management system for the construction of CMS.

Petri-Net (potentially Coloured and/or Timed Petri-Net) [6] notation is being investigated in CRISTAL to model the overall production schemes, to represent this scheme to the coordinators and to facilitate the translation of the production schemes into executable code. The research area of Flow Nets [15], object-oriented extensions to Petri-Nets designed to support modelling and maintenance of scheduling tasks associated with large scale industrial manufacturing systems, may be investigated as a mechanism for providing further production control facilities.

## 8 Present Status and Conclusions

CRISTAL development was initiated in early 1996 at CERN and a prototype data capture tool was delivered using O2 and the WWW to allow existing aspects of ECAL construction to be incorporated in a object database. This prototype demonstrated the use of the WWW with O2, allowed the capture of histograms holding such data as transmission spectra and can be viewed via Netscape (version 2.0 or above) by visiting http://hpcord02.cern.ch/cristal/main.html.

The second spiral of prototyping and technology evaluation in CRISTAL was initiated during the summer of 1996 and aims to deliver a further prototype, based on the ECAL testing and construction programme in CMS, to formulate the coupling between operators, instruments and CRISTAL, to demonstrate the implementation of a set of user-specific Java interfaces, to investigate the coordination of multiple distributed object databases and to initiate research in the area of distributed production control using Petri/Flow Nets. The CRISTAL prototypes are, however, being developed in a manner which is non-ECAL dependent. This then allows the results of the prototypes to be used by other CMS detectors as a basis for their engineering information management and control.

The CRISTAL project, being a form of workflow management system, requires the ability to store event histories and to handle dynamic schema evolution in the object database. Given that the OODBMS selected for the CRISTAL prototype, O2, cannot fully handle dynamic schema evolution, we have introduced the concept of meta-modelling in CRISTAL to by-pass these schema evolution problems. The CRISTAL project has already shown the viability and importance of adopting an object-oriented approach to the development of software whose lifetime needs to be considerable. It continues to investigate the use of PSS-05 with OO techniques and to provide a test-bed for emerging OO tools such as O2, Orbix, Java, OrbixWeb and STP.

## 9 Acknowledgements

The authors take this opportunity to acknowledge the support of their home institutes and to thank all those involved in the continuing CRISTAL effort. In particular, the continuing support of P Lecoq and J-L Faure is greatly appreciated.